\documentclass[reprint,amsmath,amssymb,aps,prb,showpacs,showkeys]{revtex4-1}
\usepackage{graphicx}
\usepackage{dcolumn}
\usepackage{bm}
\usepackage{natbib}
\usepackage{color}
\usepackage{soul,xcolor}
\usepackage{amsmath}
\usepackage{bm}   
\usepackage[T1]{fontenc}

\def \pst {Pb$_{1-x}$Sn$_x$Te}
\def \pss {Pb$_{1-x}$Sn$_x$Se}

\begin{document}
\bibliographystyle{plainnat}
\setstcolor{red}
\title{
Reconstruction, rumpling, and Dirac states at the (001) surface 
of a topological crystalline insulator \pss }

\author{ A.~{\L}usakowski}
\affiliation{Institute of Physics, Polish 
Academy of Sciences, Al. 
Lotnik\'{o}w 32/46, PL-02-668 Warsaw, Poland}
\author{P. Bogus{\l}awski}
\affiliation{Institute of Physics, Polish 
Academy of Sciences, Al. 
Lotnik\'{o}w 
32/46, PL-02-668 Warsaw, Poland}
\author{T. Story}
\affiliation{Institute of Physics, Polish 
Academy of Sciences, Al. 
Lotnik\'{o}w
32/46, PL-02-668 Warsaw, Poland\\
and\\ International Research Centre MagTop, Institute of 
Physics, PAS, Al. Lotnik\'ow 32/46, PL-02-668,
 Warsaw, Poland
}

\begin{abstract}%

Equilibrium atomic configuration and electronic structure of the (001) 
surface of IV-VI semiconductors PbTe, PbSe, SnTe and SnSe, is studied 
using the density functional theory (DFT) methods. At surfaces of all 
those compounds, the displacements of ions from their perfect 
lattice sites reveal two features characteristic of the rock salt 
crystals. 
First, the ionic displacements occur only along the direction 
perpendicular to the surface, and they exhibit the rumpling effect, 
$i.e.$, the vertical shifts of cations and anions differ. Second, the 
interlayer spacing of the first few monolayers at the surface 
oscillates. Our results are in good  agreement with the previous X-ray 
experimental data and theoretical  results where available. They also 
are consistent with the presence  of two \{110\} mirror planes at the 
(001) surface of the rock salt. One the other hand, experiments 
preformed for the topological \pss\ alloy indicate breaking of the 
mirror symmetry due to a large 0.3 \AA\ relative displacement of the 
cation and anion sublattices at the surface, which induces the 
opening of the gap of the Dirac cones. Our results for \pss\, including 
the simulated STM images, are in contradiction with these findings, 
since surface reconstructions with broken symmetry are never the 
ground state configurations. The impact of the theoretically 
determined surface configurations and of the chemical disorder on 
the surface states is analyzed.

\end{abstract}

\maketitle

\section{Introduction} 

Various propositions envisioned for applications of topological materials rely 
on control of their electronic structure. This issue is motivating a 
considerable ongoing experimental and theoretical research. As it was 
demonstrated, properties of topological electron states can be tuned by gate 
voltage, crystal strain, magnetic field, or magnetization. This includes in 
particular the opening of the energy gap in the otherwise metallic Dirac-like 
spectrum. 

In the case of the IV-VI narrow gap semiconductors \pst\ and \pss\ studied 
here, it is possible to observe a transition from a normal insulator (NI) to a 
topological crystalline insulator (TCI) phase  by changing chemical 
composition, external pressure, or temperature. Those materials crystallize 
in the rock salt ({\it rs}) structure.\cite{nimtz,khoklov} The key physical factor 
responsible for the existence of the nontrivial topological phase is not the 
time reversal symmetry, as is the case of well known topological insulators 
(Bi,Sb)$_2$(Te,Se)$_3$, but the mirror symmetry with respect  to the \{110\} 
crystallographic planes [\onlinecite{hsieh, dziawa,tanaka,xu}]. This symmetry 
warrants the existence of zero gap Dirac-like states on certain high-symmetry 
crystal facets of TCI  crystals, like (001), (111) or (110). 

Scanning tunneling microscopy and spectroscopy (STM/STS) and Landau 
levels spectroscopy (LLS) experiments suggested that at the (001) surface of 
\pss\ the anion and cation sublattices are displaced with respect to each 
other along the [110] direction.\cite{okada,zeljkovic} Such a displacement 
preserves the mirror symmetry with respect to the (110), but breaks the 
symmetry with respect to the ($\bar{1}$10) plane. This symmetry breaking 
gives rise to the opening of the gap in two out of total four valleys of surface 
Dirac TCI states located near the $\bar{X}$-points of the surface Brillouin 
zone (projections of four bulk $L$-points) on $\bar{\Gamma} - \bar{X}$ lines. 
Recently, such a symmetry breaking was also observed in \pst\ 
[\onlinecite{nishi}]. The results of STS/LLS studies in magnetic field 
[\onlinecite{okada}] confirmed the conclusion that two kinds of surface 
states coexist, those with a vanishing energy gap and those with a finite one. 
In the subsequent paper [\onlinecite{zeljkovic}] the STM images directly 
showed the relative displacement of sublattices of about 0.3~\AA. In both 
papers, only the low temperature measurements were reported. This 
problem was also studied experimentally in Ref. [\onlinecite{wojek}] by 
angle-resolved photoemission spectroscopy (ARPES), a technique able to 
selectively study different valleys in the $k$-space. Importantly, the analysis 
included samples with different Sn concentrations, and was performed at 
different temperatures. In agreement with Refs. \onlinecite{okada} and 
\onlinecite{zeljkovic}, at low temperatures two kinds of surface states were 
observed -- gapless and with the energy gap of the order of 25 meV 
[\onlinecite{wojek}]. The surface gap decreased with the increasing 
temperature and, depending on $x$, it disappeared in the temperature 
range 100 -- 200 K, as expected from the topological $T-x$ phase diagram of 
\pss. \cite{wojek1} In the papers\cite{okada, zeljkovic,wojek} the mechanism 
of symmetry breaking at the surface of bulk crystals with the perfect cubic 
symmetry was not proposed. Instead, the effect was tentatively ascribed to 
the tendency of the IV-VI crystals to the transition from the cubic to the 
rhombohedral phase observed in tellurides SnTe, GeTe, and 
\pst.\cite{nimtz,khoklov} 

The surface reconstruction proposed in [\onlinecite{zeljkovic}] is
unexpected. This is because previous theoretical and experimental
works on \{001\} surfaces of compounds crystallizing in the rock salt
structure \cite{verwey, lazarides, sawada, deringer, deringer1, ma,
satta} revealed that in all cases considerable displacements from the
ideal bulk sites take place, but their character is in qualitative
disagreement with that found in [\onlinecite{zeljkovic}]. In particular, 
three features universally characterize surface equilibrium
geometries\cite{verwey, lazarides, sawada, deringer, deringer1, ma,
satta}. First, atomic displacements occur in the $z$-direction 
(perpendicular to the surface) only,
thus both the ideal {\it rs} surface periodicity and symmetry are
maintained. Second, the top surface layer is not ideal flat, because
cations and anions shift by different amounts in the $z$ direction;
this effect is referred to as rumpling.
Third, the average interlayer spacings relative to the ideal bulk
geometry are modified, and the distance between the first and the
second layer is reduced while that between the second and the third
layer is increased compared to the ideal bulk value. Such an
oscillatory behavior occurs for the first few subsurface atomic
layers. No (110) mirror plane symmetry breaking reconstructions were
found.
Sawada and Nakamura\cite{sawada} summarized early works, and
used the Verwey model\cite{verwey} to show that the relaxations
are largely determined by electrostatics. In fact, responsible for the
effects above are both long range electric fields and different atomic
polarizabilities of cations and anions. 

Results for \pst\ are contraditory as well. Yan {\it et al.}
\cite{yan} studied the band structure  of \pst (111) overlayers in the
full  
composition range. The Dirac states were seen by ARPES also for the TCI 
SnTe, and neither band gap nor crystal structure anomalies were noticed. On 
the other hand, recently a symmetry breaking analogous to that in \pss\ was 
also observed [\onlinecite{nishi}].

Here, we theoretically study the ground state atomic configurations  and the 
electronic structure of the (001) surfaces of some  semiconductors from the 
IV-VI family. Of particular interest are the  possible mechanisms of symmetry 
breaking and the gap opening in  surface states.  The paper is organized as 
follows. In Section II we  present details of the ab initio calculations, together 
with the method  of obtaining the Tight Binding Approximation (TBA) 
Hamiltonian from  the {\it ab initio} results. Section III is devoted to the 
geometric  optimization of the considered systems. Our results are in full 
accord  with the previous findings\cite{verwey, lazarides, sawada, deringer,  
deringer1, ma, satta} regarding the features of surface geometry of  the {\it 
rs} crystals listed above, but do not confirm the surface reconstruction 
proposed in  Ref. [\onlinecite{zeljkovic}]. An additional insight is obtained 
from   our simple simulations of the STM images. Finally, electron dispersion 
relations of the surface states are provided, where we also  study the 
influence of both surface deformations and chemical  disorder on the band 
gap. This allows to determine how a realistic atomic-scale surface 
morphology of the crystal facets hosting topological states influences 
electronic spectrum. We identify the conditions under which the Dirac-cone 
like metallic spectrum can be shifted in the $k$-space along specific 
directions, or can exhibit about 1-10 meV gap opening effect relevant, for 
e.g. FIR and THz applications. Section IV concludes the paper.

\section{Methods of calculations} 

\begin{figure} 
\includegraphics[width=\linewidth]{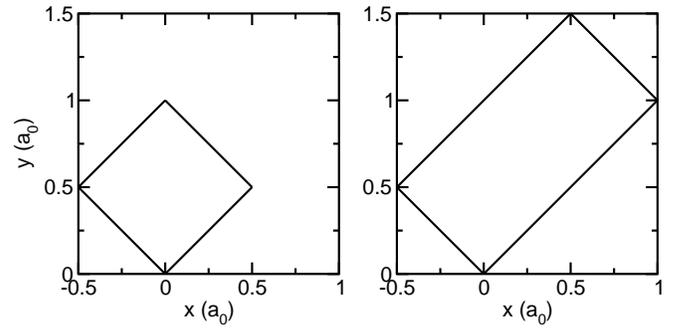} 
\caption{\label{fig1} Surface unit cells for (a) SC1 and SC2 (b) 
supercells, see text.} 
\end{figure} 

Properties of layers are analyzed using slabs with the identical top 
and bottom layers, and thus with the odd  number of monolayers. 
The slabs are separated by 10 \AA\ of vacuum.
For pure compounds 
the  supercell SC1 is defined by the vectors ${\bm  
a_1}=a_0(\frac{1}{2},\frac{1}{2},0)$, ${\bm  a_2}=a_0(-\frac{1}{2}, 
\frac{1}{2},0)$ and ${\bm a_3}=a_0(0,0,n_h+\frac{1}{2})$, where
$2n_h+1$ is the number of monolayers in the slab.  The 
surface unit cell contains one cation and one anion.
For mixed 
crystals, when we want to take into account the symmetry breaking 
due  to chemical disorder, we use larger supercells SC2 based on 
vectors ${\bm  a_1}=a_0(1,1,0)$,  ${\bm  a_2}=a_0(-\frac{1}{2}, 
\frac{1}{2},0)$ and  ${\bm  a_3} = a_0(0,0,n_h +\frac{1}{2})$. The 
supercell SC2 consists of two  supercells SC1, what enables an easy  
presentation of dispersion relations. The surface unit cells of SC1 and 
SC2 are shown in Fig.~\ref{fig1}.   

The DFT calculations are performed using OpenMX package. 
\cite{openmx} They are based on the local density approximation 
\cite{CA} and the exchange-correlation functional of Ref. 
[\onlinecite{PZ}]. Summations over the Brillouin Zone are 
performed applying a $4\times 4\times 1$ mesh, with the convergence 
check using a $8\times 8\times 1$ mesh for SC1, 
and $4\times 8\times 1$ for SC2. Atomic pseudopotentials 
for Pb and Te were described in Ref. [\onlinecite{lusakowski1}], 
and those for Sn and Se were taken from OpenMX.
Atomic relaxations in the first five surface layers are taken into 
account. The calculations were stopped when the forces acting on 
atoms were smaller than $10^{-4}$ Ha/Bohr, and total energy was 
converged to within $10^{-6}$ Ha. 

The calculated equilibrium {\it rs} lattice parameters $a_0$ are 
6.42~\AA\ for PbTe, 6.40~\AA\ for SnTe, and 6.14~\AA\ for PbSe.
These values are slightly different from the respective experimental 
ones which are 6.46~\AA, 6.30~\AA, and 6.12~\AA, respectively.
The experimental value for {\it rs}-SnSe 
is not known since SnSe in the {\it rs} structure is unstable, our
calculations result in 6.10~\AA\ for {\it rs} SnSe.
Using the theoretical lattice parameters we optimize 
surface geometry for 11 and 23 monolayer (ML) thick slabs, 
and the equilibrium atomic positions are practically identical.

\begin{figure}[t]
\includegraphics[width=\linewidth]{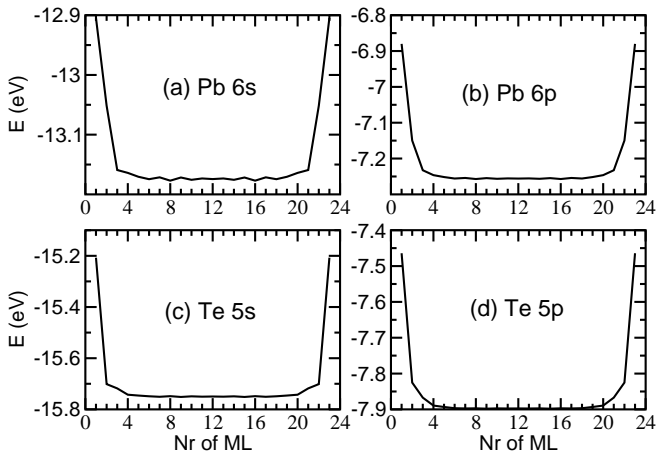} 
\caption{\label{fig2} Calculated dependence of the TBA parameters, 
namely the energies of the 6$s$(Pb), 6$p$(Pb), 5$s$(Te), and 5$p$(Te)
 orbitals, on the atomic position in the 23-ML slab. } 
\end{figure} 

Reliable calculations of the dispersion relations of surface states 
require  
sufficiently thick slabs. Otherwise, the relevant surface states 
localized at the top 
and at the bottom surfaces interact, which leads to a hybridization 
driven  
opening of energy gaps in the spectrum of the surface states. Our 
convergence checks indicate that the coupling between the two 
surfaces is negligibly small for slabs of the thickness of about 150 
ML, in agreement with Ref. [\onlinecite{yan}]. 
This corresponds to about 600 atoms in the supercell, and 
thus the {\em ab initio} calculations, although in principle possible, 
are nonpractical.  

An efficient alternative to apply is the TBA approach based on the 
LDA results. This requires the knowledge of TBA parameters, and 
they are provided by the OpenMX code. Here, we use the TBA 
parameters determined by {\it ab initio} calculations for 23 ML 
thick slabs. In the slab geometry, an issue to solve 
is that the TBA parameters for a given atomic species ($e.g.$, Pb in 
PbTe) depend on the ion's position in the slab. Indeed, because the 
local coordination of a Pb ion at the PbTe surface is not the same as 
in bulk, the parameters of Pb near the surface differ from those in 
bulk, and this difference depends on the distance from the slab's 
surfaces. 
This also holds in the case of the interatomic TBA parameters. According 
to our results, significant variations of the parameters are limited to 
4 ML nearest to surfaces. This important fact was typically ignored in 
the literature. 

\begin{figure} [b]
\includegraphics[width=\linewidth]{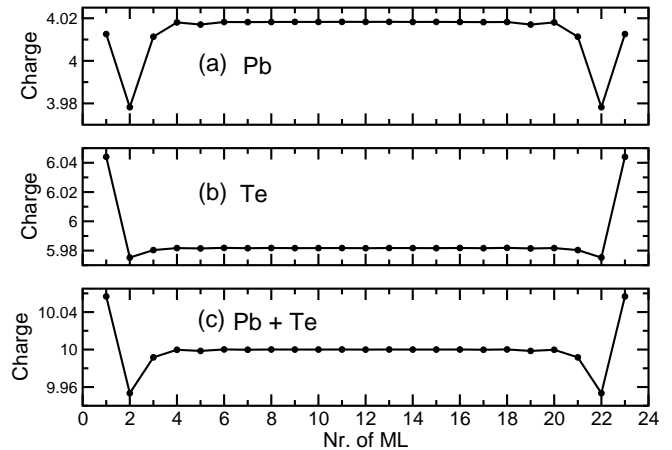} 
\caption{\label{fig3} Dependencies of Mulliken charges on the 
atom positions in the 23 ML slab: (a) Pb, (b) Te, and (c) the sum of (a) 
and (b).} 
\end{figure} 

\begin{figure} 
\includegraphics[width=\linewidth]{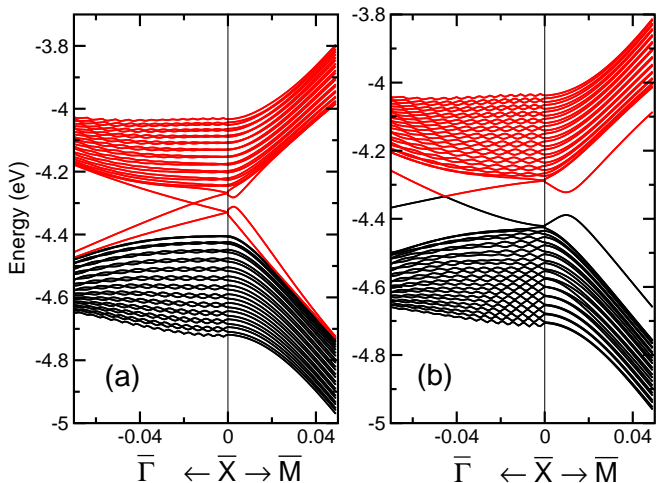} 
\caption{\label{fig4} (color online)  
Dispersion relations along $\bar{X}\rightarrow\bar{\Gamma}$ and 
$\bar{X}\rightarrow\bar{M}$ directions of the 25 highest valence 
(black lines) and the 25 lowest conduction (red lines) bands for 123-
monolayer thick slab of PbTe with the lattice parameter 
$a_0$=6.20~\AA. (a) The TBA parameters are independent of the 
position in the slab, and (b) the TBA parameters are position 
depended, see text. The wavevector on the $x$ axis along both the 
$\bar{X}\rightarrow \bar{\Gamma}$ and $\bar{X}\rightarrow 
\bar{M}$ directions is in the units of $2\sqrt{2}\pi/a_0$.} 
\end{figure}

As an example, in Fig.~\ref{fig2} we show the dependence of 
the TBA 
energies of $s$ and $p$ orbitals of both Pb and Te ions on 
their positions in the PbTe slab.  Energy differences between the values 
in 
bulk and at the outermost surface layer are 0.3 eV for the
Pb orbitals, and somewhat larger, about 0.5 eV, for Te orbitals. 
The dependencies of the Mulliken charge on atomic 
positions in the slab, shown in Fig.~\ref{fig3}, reveal the 
same 
feature: their significant variations are limited to the first four 
layers. 
The difference in the values for the first ML and at the slab center is 
not 
very large, being of the order of 2 \%. Similar differences are found 
for other TBA parameters, and for the remaining crystals. Importantly 
however, these variations have a critical impact on the energy 
dispersion relations and the character of the surface states. 
This is illustrated in Fig. \ref{fig4}, which 
presents the dispersion relations for a 123 ML thick slab of PbTe 
along the $\bar{\Gamma}\rightarrow\bar{X}$ and 
$\bar{X}\rightarrow\bar{M}$ directions obtained within the TBA. 
Only the highest 25 valence states and the lowest 25 conduction 
states are displayed. The bulk lattice parameter $a_0=6.20$ \AA\ is 
assumed, for which PbTe is in the nontrivial topological 
state. \cite{lusakowski2} The results of the panel (a) are obtained 
with all TBA parameters equal to the TBA parameters of bulk PbTe. In 
the case of the panel (b), the TBA parameters are position dependent 
and determined according to the procedure described below. We 
expect that the highest occupied levels should be the surface states, and 
indeed such a situation is observed with a well defined Dirac cross, 
see the panel (b). In contrast, in the case (a) there is a number of 
bands above the energy of the highest occupied band, which 
certainly cannot be classified as conduction bands. We conclude 
therefore, and this is one of important messages of our work, that 
the position dependence of the TBA parameters close to the surface 
must be taken into account. 
 
The position-dependent TBA parameters for thick layers are 
obtained as follows. We begin with {\em ab initio} calculations for 
an easy to handle 23 ML slab with equivalent top and bottom 
surfaces. The slab is partitioned into 3 regions, the bottom (I), the 
internal (II) and the top (III) one, containing 6, 10 and 7 ML, 
respectively. The corresponding TBA parameters are given in the 
output of {\em ab initio} calculations. The assumed thickness of 
the outer regions I and III guarantees that all surface effects are 
contained therein, and thus the internal region II reproduces 
correctly the bulk. In particular, in our TBA method the 
interatomic couplings extend to the third neighbor of a given 
atom, and the atomic relaxations along with the effects of the 
non-bulk coordinations at the surface extend to the first four ML (see 
Figs \ref{fig2} and \ref{fig3}). 
The TBA parameters for the internal 
region II are those for bulk crystals. Also
for thicker slabs three regions are considered. The first and 
the last regions contain 6 and 7 ML, respectively, and the internal 
region is a certain number of repetitions of the region II of 23 ML 
slab. For these three regions the TBA parameters are taken from I, 
II, and III regions of a 23 thick ML slab, respectively. 

\section{Atomic configurations  and electronic structure of (001) 
surfaces } 

\subsection{Ground state atomic configurations}

Atomic configuration at a (001) surface of a rock salt crystal can be 
defined by two parameters,\cite{satta} namely the average $z$ 
coordinate of the $i$-th layer $z(i) = (z_{cat}(i) + z_{an}(i))/2$, and 
the rumpling parameter of that layer $r(i) = (z_{an}(i) - 
z_{cat}(i))/d_0)$. Here, $d_0$ is the equilibrium bond length, and 
the $z$-coordinate of cations (anions) from the $i$-th layer with 
respect to perfect positions is denoted by $z_{cat}(i)$ 
($z_{an}(i)$), with $i=1$ being the surface. The change in the 
interlayer spacing between the layers $i$ and $i+1$ is 
$\delta_{i,i+1} = (z(i+1) - z(i))/d_0$. To find the ground state atomic 
configurations, three initial 
geometries are considered. In the first one, all atoms are at the 
ideal rock salt sites. In the remaining cases, the initial 
displacements for Te atoms are limited to the surface layer only, 
and amount to (0.3, 0.3, 0)~\AA\ and (0.4, 0.2, -0.1)~\AA, 
respectively. These choices are meant to initiate the surface 
configuration suggested in Ref. [\onlinecite{zeljkovic}], which 
breaks the surface symmetry. 

\begin{figure} 
\includegraphics[width=\linewidth]{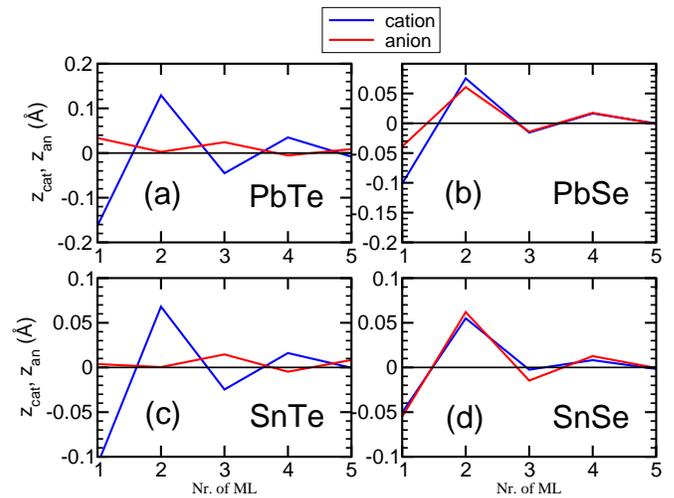} 
\caption{\label{fig5}  (color online)   
Displacements of cations and anions from the ideal rock salt sites in 
the first 5 monolayers for (a) PbTe, (b) PbSe, (c) SnTe, and (d) SnSe.} 
\end{figure}

\begin{figure}[t] 
\includegraphics[width=\linewidth]{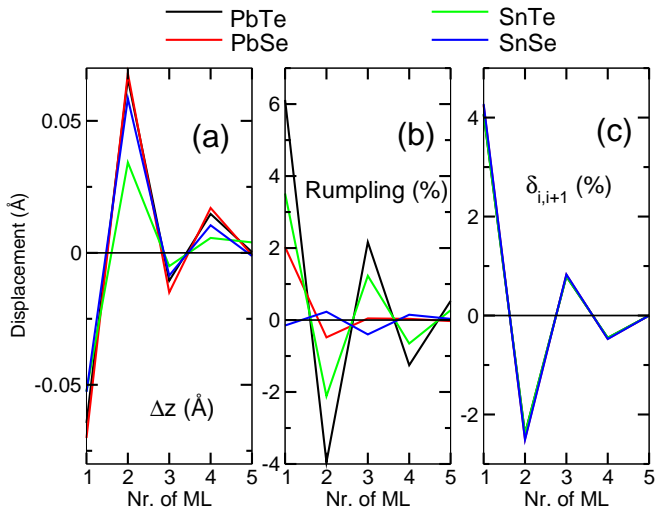} 
\caption{\label{fig6}  (color online)  
Changes of (a) the average $z$ coordinates of the monolayers, 
(b) rumpling $r$, and (c) of the interlayer spacings $\delta$ 
for first 5 monolayers of PbTe, PbSe, SnTe and SnSe. Note that the
calculated $\delta$s are practically identical for all  the considered
crystals. 
} 
\end{figure}

We first consider ionic relaxations perpendicular to the surface. 
The calculated final displacements $z(i)$ along the $z$ axis do not 
depend on initial configuration, and are shown in Fig.~\ref{fig5} 
for all the considered compounds. Next, both the rumpling and 
the interlayer spacing are presented in detail in Fig.~\ref{fig6}. For 
PbTe and PbSe, these parameters were calculated 
previously.\cite{lazarides,satta,ma} There are some differences 
between our values and those presented in the literature but the 
qualitative picture is the same. The results for SnSe and SnTe are 
new to our best knowledge. 

The PbTe(001) surface was experimentally investigated in detail. 
LEED measurements performed in Ref. [\onlinecite{lazarides}] 
revealed a large rumpling, $r(1)=6.8 \%$, which corresponds to 
displacements of about 0.2 \AA. Next, the measured interlayer 
spacings exhibit an oscillatory behavior, with $\delta_{1,2}=4\%$ 
and $\delta_{2,3}=-2\%$. Our results shown in Fig. \ref{fig6} are 
in a good agreement with these data: the rumpling of the first 
layer is $r(1)\approx 6\%$, and the changes in interlayer spacings 
are $\delta_{1,2}\approx 4\%$ and $\delta_{2,3}\approx -2\%$. 
Previous first principles calculations for PbTe(001) \cite{lazarides, 
satta, ma} found the rumpling somewhat lower than that 
observed, but they confirmed the outward (inward) shift of Te (Pb) 
ions. Next, the calculated interlayer spacings exhibited an oscillatory 
behavior with values close to those measured. 

\begin{figure}[b]
\includegraphics[width=\linewidth]{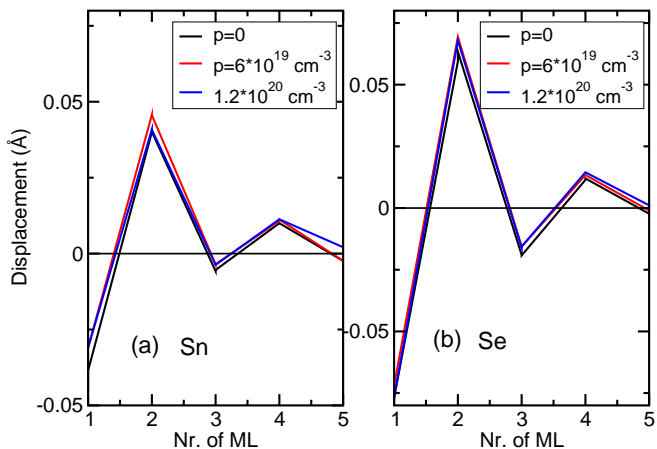} 
\caption{\label{fig7}  (color online)  
Displacements of atoms in the first monolayers of SnSe along the [001]
direction with respect to the positions 
in the perfect rock salt lattice after the geometric optimization 
as the function of the hole concentration $p$. 
The results for Sn and Se are given in Fig. (a) and (b), respectively.} 
\end{figure}

Rumpling in the PbX series depends on the anion. Indeed, in 
contrast to PbTe(001), a very small rumpling at the PbS(001) surface 
was observed \cite{kendelewicz}. Theoretical results of Refs. 
\cite{satta, ma, deringer} are in a good agreement with experiment. 
Interestingly, rumpling at the PbS(001) and PbTe(001) surfaces is 
opposite, since in the former case the S anions exhibit a small 
$inward$ displacement. According to Fig. \ref{fig5} and Fig. \ref{fig6}, 
relaxations at 
PbSe(001) are intermediate between those for PbS(001) and 
PbTe(001). In particular, the rumpling at the PbSe (001) is smaller 
than for PbTe as it amounts to 2 \%. This confirms the results of 
Ref.~[\onlinecite{deringer}]. The interlayer spacing exhibits the 
typical oscillatory behavior.  

The results obtained for {\it rs}-SnSe and {\it rs}-SnTe are presented in 
Figs. \ref{fig5} and \ref{fig6}, respectively. 
As we see, the $z$-displacements of the  
consecutive layers are similar for all systems, and the interlayer 
distances always oscillate. Also the rumpling oscillates, changing the 
sign for adjacent layers, but it is more pronounced in the tellurides 
than in the selenides. We recall here that the {\it rs} phase of SnSe is 
only metastable, but nevertheless surface properties of 
{\it rs}-SnSe(001) 
are expected and found to be similar to those of other members of 
the {\it rs}-IV-VI family. 

Equilibrium atomic configurations can depend on the presence of free 
carriers. Indeed, it is well known that Sn-based IV-VI compounds typically 
are highly $p$-type due to the presence of high concentrations of 
electrically active cation vacancies. This is also the case of crystals 
investigated in Ref. [\onlinecite{zeljkovic}]. There are two possible 
consequences of this fact. First, charge transfer processes taking place at 
the surface, see Fig. 3, can be screened and/or affected by the presence of 
free carriers in the layer. This in turn can influence the surface geometry. 
Second, bulk SnTe assumes the NaCl structure at higher temperatures, and 
with the decreasing temperature a transition to the rhombohedral phase 
takes place. However, this transition is supressed by concentrations of 
holes higher than 1.5 $\cdot$ 10$^{20}$ cm$^{?3}$ [\onlinecite{kobayashi}]. 
(This transition also becomes suppressed with the increasing content of Pb 
in \pst\.) Accordingly, the recent experiment of Wei {\it et
  al.}\cite{wei}
suggested that strongly anomalous transport features observed in 
PbTe/SnTe heterostructures originate in the free-carrier driven symmetry 
breaking in SnTe layers caused by the above mentioned varying hole 
densities. 

The results above motivated us to analyze the impact of free carriers on 
atomic displacements. OpenMX package allows performing calculations for 
nonzero concentrations of free carriers. The calculations were done for 
free carrier concentrations between 10$^{20}$\ cm$^{-3}$ of holes to 
10$^{20}$\ cm$^{-3}$ of electrons, and the results are shown in Fig. 
\ref{fig7}. While the relaxation energies ($i.e.$, the energy gains induced 
by the relaxation from the ideal to the equilibrium configuration) are 
dependent on carrier concentration to some extent, the final geometries 
are highly non-sensitive to the presence of free carriers. In all cases, the 
ionic displacements in the (001) plane are negligibly small. The presence of 
1.2$\cdot$10$^{20}$\ cm$^{-3}$ electrons induces ionic displacements of 
about 0.01 \AA. The same amount of holes induces a similar change, but of 
opposite sign. 

We now turn to the displacements in the $(x, y)$ plane. When the 
ideal {\it rs} configuration is used as input, the final ionic 
displacements in the $(x, y)$ plane always vanish for all the 
considered crystals. Using the symmetry-breaking initial 
configurations leads to final configuration with small 
displacements of about |0.03|~\AA. 
We ascribe this result to the finite accuracy of our calculations, 
since in all cases total energy of the symmetry breaking 
configurations is higher by about 0.5 meV/(surface atom) 
than that of the symmetric ground state. 

Concluding, our results do not confirm the presence of a 
symmetry breaking in the $(x,y)$ plane at the surface, which were 
proposed in Ref. [\onlinecite{zeljkovic}] based on the STM 
measurements. In fact, our study indicates instability of those 
configurations. Also the presence of free carriers can be ruled 
out as a cause of the surface reconstruction reported in Ref. 
[\onlinecite{zeljkovic}]. Independent experiments, in 
particular STM, are needed to clarify the situation.

\subsection{Simulation of STM images}

\begin{figure} 
\includegraphics[width=\linewidth]{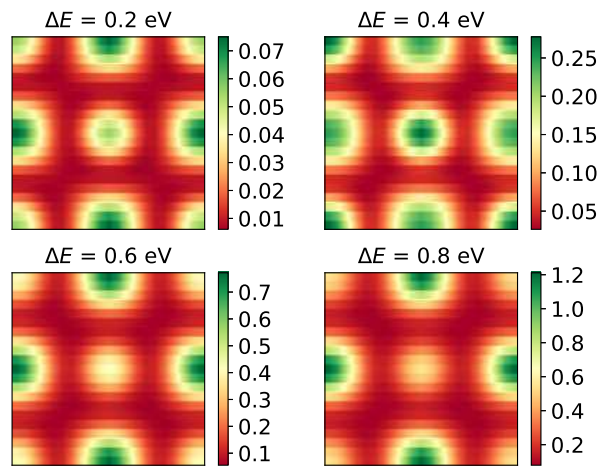} 
\caption{\label{fig8}  (color online)  
Integrated local density of states $n(\bm r,E)$ at 1.5 \AA\ from the 
PbSe(001) surface calculated for the varying energy window $\Delta 
E$ counted from the top of the valence band, see text. The figures
display the surface unit cell of dimension $a_0\times a_0$. Pb cations 
are located at the center and at the corners of the unit cell, 
and Se anions are at the middle of its edges. } 
\end{figure}

Conclusions reported in Ref. [8] regarding the surface geometry are 
based on the STM measurements. In those measurements the local 
density of states near the surface plays the main role.\cite{tersoff} To 
make a link with experiment we simulate, in a simplified way, the STM 
images by calculating the integrated local density of states (ILDOS) 
$n(\bm r, E)$ at the distance 1.5 \AA\ from the surface according to  
\begin{equation} 
n({\bm r}, E)=\sum_{E<\epsilon_{nk}<0}|\psi_{nk}({\bm r})|^2, 
\end{equation} 
\noindent 
where the summation is performed over all electron states with 
energies $\epsilon_{nk}$ belonging to the energy window between 
$E$ and the top of the valence band (taken as the zero energy). The 
current in STM measurements is proportional to the ILDOS defined 
above. In this way, different voltages applied in STM measurements 
are modeled by different energy windows $\Delta E = -E$. A negative 
value of 
$E$ implies that we sum over the occupied valence states, which 
corresponds to the voltage polarization used in Ref. [8]. The results 
obtained for PbSe(001) are displayed in Fig. \ref{fig8} which 
shows the voltage dependence of STM images. 
Qualitatively, the figure reproduces the effect presented in Fig. S3 of 
Ref. [8], and in particular the fact that different atoms are dominant 
under different STM voltages. In general the anions are more visible
because, due to rumpling, the cations are placed deeper in the
crystal. However, as expected, the STM images  
of the symmetric PbSe(001) surface are symmetric as well, and the 
maxima of the ILDOS $n(r,E)$ are above the ions. 

\begin{figure} 
\includegraphics[width=\linewidth]{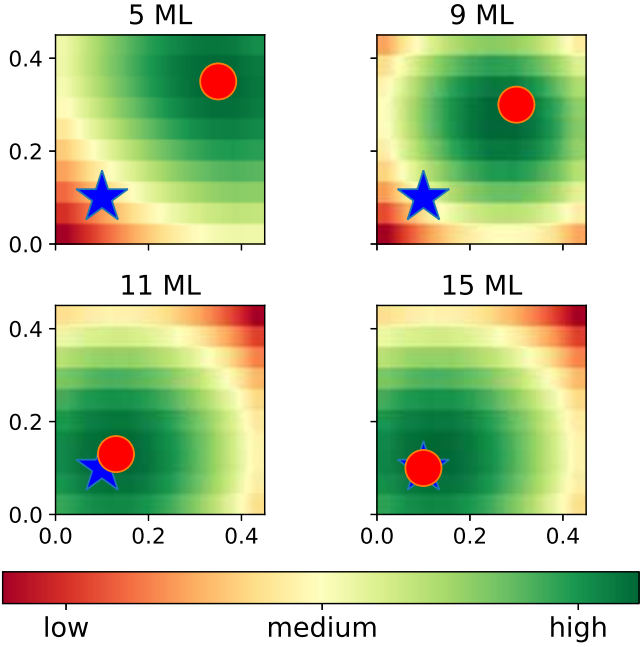} 
\caption{\label{fig9}  (color online)  
ILDOS at the distance 1.5 \AA\ from the PbSe(001) surface near the 
displaced Se atom for $E =-0.6$ eV. $\Delta x$ and $\Delta y$ describe
distances relative to the position of Se atom in the perfect {\it rs} 
structure. 
The  actual position of Se is denoted by the blue star, and 
the maximum of 
the calculated electron density occurs at the red circle. } 
\end{figure}

The simulated STM images change when the surface symmetry is broken, 
and a relative displacement of the cation and anion sublattices at the 
surface is assumed. In this case, calculations performed for several 
surface 
configurations reveal an interesting effect illustrated in Fig. 
\ref{fig9}. The 
images are obtained assuming a lateral displacement of Se in the first ML 
by 
(0.1, 0.1)~\AA\, and the Se position is shown by a star. On the other 
hand, 
the calculated maxima of the electron density occurs at approximately 
(0.3, 
0.3)~\AA\ in the case of the very thin slabs of 5 and 9 ML. This result 
shows 
that the experimental STM images do not always directly reflect location 
of 
atoms at the surface. However, for thicker slabs the effect  disappears.  
Because in Ref.~[\onlinecite{zeljkovic}] the surface of  bulk crystal was 
studied, STM maxima reflect real positions of atoms.  
(We mention here that the sensitivity of the PbTe band gap on the slab 
thickness discussed in Ref.~[\onlinecite{bessanezi}] showed 
pronounced   irregularities for a few monolayer slabs. Also, in general 
atomic configurations suggested by the STM images do not necessarily 
reflect the actual ones. For example, at the Si(001) surface the apparent 
tilt 
of Si dimers is   different for positive and negative bias 
\cite{boguslawski}. )

\subsection{Dispersion relations of surface states. 
Influence of rumpling}

Surface states of the considered systems are affected both by 
rumpling and by chemical disorder present in alloys. The latter 
destroys the reflection symmetry with respect to the \{110\} planes. 
We focus on the two surface states present in the band gap because 
we are interested mainly in the position of the Dirac cross and the 
energy gap in the vicinity of the $\bar{X}$ point. Dispersion relations 
for 133 and 173 ML thick slabs and for [110] and [$\bar{1}$10)] 
directions 
are identical.

\begin{figure}[t]
\includegraphics[width=\linewidth]{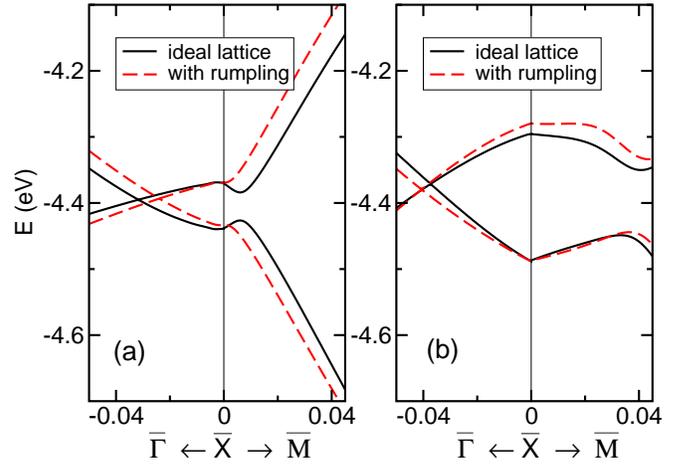} 
\caption{\label{fig10}  (color online)  
The influence of rumpling on the energy bands for 
(a) PbTe with the lattice parameter $a_0$=6.24 \AA, when PbTe is in 
the TCI phase, and (b) {\it rs}-SnSe with $a_0=6.105$~\AA. 
In both panels only the two surface states present in the band gap 
are shown. 
The wavevector on the $x$ axis along both the $\bar{X}\rightarrow 
\bar{\Gamma}$ and $\bar{X}\rightarrow \bar{M}$ directions is in the 
units of $2\sqrt{2}\pi/a_0$.}
\end{figure}

Figure \ref{fig10} compares dispersion relations for surfaces of 
both PbTe and {\it rs}-SnSe with and without rumpling. The calculations 
for {\it rs}-SnSe are done for comparison only because, as we already 
mentioned, the {\it rs} phase of SnSe is not stable. 
In the case of PbTe, the assumed lattice parameter is 6.24~\AA\, for 
which the 
compound is in the TCI phase. We see that the rumpling does not 
lead to opening of the gap at the Dirac cross because the reflection
symmetry with respect to the \{110\} planes is not broken. 
There are  two effects induced by rumpling in the vicinity of the 
$\bar{X}$ point, namely a displacement of the Dirac cross 
and a change of the energy gap on the $\bar{X}$-$\bar{M}$ direction. 
Both effects are very small. 
For {\it rs}-SnSe we also observe these two effects, however, 
compared to PbTe, the shift of the Dirac cross is of opposite 
sign and the energy gap becomes larger. 
Similar to the case of PbTe, both effects are small.

\subsection{Bulk chemical disorder}

In a pure compound like PbSe, transition from the trivial NI to the TCI 
phase can be induced by hydrostatic pressure. Since hydrostatic 
pressure does not change the crystal symmetry, the states at the four 
$L$ points in the Brillouin zone are degenerate for all values of the 
lattice constant. Consequently, the transition is sharp, and the band 
gap changes sign at a well defined value of $a_0$.  

In alloys, chemical disorder makes the NI-TCI transition more complex. 
This is because the disorder splits degenerate bands (or it broadens 
electronic states), and as a consequence the pressure-induced NI-TCI 
transition is smeared out, $i.e$, it occurs within a finite window of 
hydrostatic pressures. In an analogous way, the NI-TCI transition can 
be induced by changing the chemical composition $x$ of an alloy, but 
again the band smearing results in a finite composition window in 
which the band gap vanishes. This was discussed for bulk crystals  in
Refs \onlinecite{lusakowski2} and \onlinecite{zunger2}.  Moreover, it
was shown in Ref. [\onlinecite{zunger2}]  
that in the transition region the alloy is in the Weyl semimetal (WSM) 
phase. Our calculations show that \pss\ is in the WSM phase for $0.18 < x
< 0.30$, which is close to the window $0.12 < x < 0.30$ obtained in
Ref. [\onlinecite{zunger2}]. 
\begin{figure}[b] 
\includegraphics[width=\linewidth]{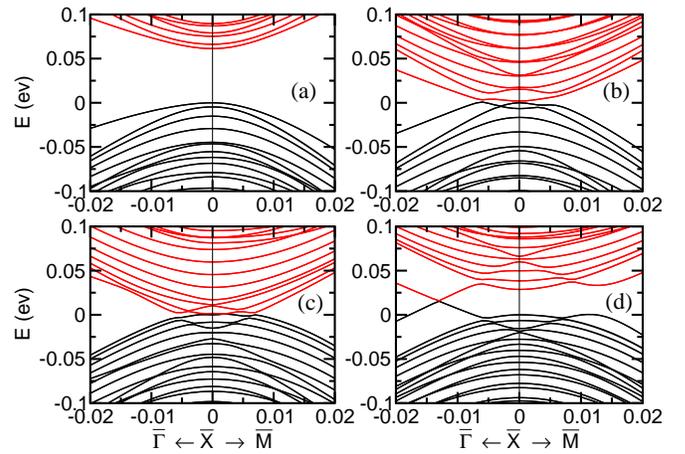} 
\caption{\label{fig11}  (color online)  Dispersion relations along the 
$\bar{\Gamma} -
  \bar{X}$ and the $\bar{X} - \bar{M}$ directions  
for \pss\ with different tin content: (a) $x$=0.15, (b) $x$=0.20, (c) 
$x$=0.25 and (d) $x$=0.30, respectively. The scales for $x$ axes are in
units of the distance between $\bar{\Gamma}$ and $\bar{X}$ points.} 
\end{figure}

In Fig.~\ref{fig11} we present dispersion relations for \pss\ 173 ML
thick slab with 
different compositions $x$ and random cation distributions in the 
supercell. For 
the lowest content of Sn, $x=0.15$ (Fig. \ref{fig11}a) the crystal is
in the NI phase and the  band gap is positive. 
Detailed analysis shows the splitting of the two lowest conduction 
and the two highest valence bands at the $\bar{X}$ point. This
is a consequence of the chemical disorder, 
and in the case of pure PbSe such a splitting is absent.
When $x=0.3$ (Fig. \ref{fig11}d) the crystal is in the TCI phase, 
and accordingly the Dirac cross is seen. 

Finally, for two intermediate concentrations $x=0.20$ and $x=0.25$
(Figs. \ref{fig11}b and \ref{fig11}c) 
the highest valence bands and the lowest conduction bands overlap 
in a finite energy interval of a few meV in the vicinity of the $\bar{X}$ 
point 
reflecting the impact of the chemical disorder. This is interpreted as 
a manifestation of the WSM bulk phase.\cite{lusakowski2,zunger2}

\section{Conclusions} 

The presence and the form of metallic helical Dirac states critically 
depend on the symmetry of not only a bulk crystal, but also of its 
surfaces. Here, we study properties of the (001) surfaces of PbTe, PbSe, 
SnTe and SnSe, as well as of the \pss\ substitutional alloy, and pay 
attention to the 
possible interplay between the topological TCI phase and the surface 
geometry. Realistic allocations of atoms in the first few monolayers at 
the (001) surface are determined by first principles calculations. 
Equilibrium geometries of all the considered surfaces are typical of the 
{\it rs} crystals family. First, the rumpling takes place, $i.e.$ cations 
and anions 
from the same atomic plane are displaced from the ideal {\it rs} sites in 
the $z$ 
direction perpendicular to the surface. In contrast, ionic displacements 
in the ($x,y)$ plane of the surface are absent, and therefore the {\it rs} 
surface symmetry is not broken. Second, close to the surface, the 
interlayer distances weakly oscillate. Those features are in agreement 
with a number of experimental data and theoretical results regarding 
surface geometries of the IV-VI family and other {\it rs} crystals. As 
these 
effects do not brake the mirror-plane symmetry of the {\it rs}(001) 
surfaces, 
their sole impact on the band structure is to shift the positions of the 
Dirac points, while opening of the band gap in the spectrum of surface 
states does not occur. In most cases, the effect of the surface 
relaxation on the band energies is small, of the order of a few meV. The 
{\it rs} symmetry is present also in our simulated STM images of the 
studied 
surfaces. On the other hand, a spontaneous symmetry breaking at the (001) 
surface, accompanied with the band gap opening (destruction of the Dirac 
cones, and acquisition of mass by surface electrons) were recently 
reported for \pss\ [\onlinecite{zeljkovic}] and \pst\ 
[\onlinecite{nishi}]. 

The above discrepancy between experiment and theory requires a comment. 
Typically, surface reconstructions stem from a specific coordination of 
surface atoms. An example is formation of dimers at \{001\} surfaces of 
zinc blende crystals. An unexpected reconstruction (which we may possibly 
deal with here) can take place because the film, assumed to be in $e.g.$ 
the {\it rs} phase, acquires a different crystal structure. Indeed, during 
the pseudomorphic growth an overlayer adopts the structure of the 
substrate. When the two structures differ, at some conditions (defined by 
the critical values of thickness, composition, or temperature) the ground 
state phase of the overlayer overcomes pseudomorphic constraints, and the 
overlayer changes its structure. Clearly, this results in a change of 
atomic configurations at the surface as well. With this respect we note 
that the equilibrium structure of SnSe and \pss\ ($x>0.4$) is
orthorhombic,\cite{szczerbakow, neupane} in which phase 
the (001) surface can acquire the reconstruction pattern claimed in  
[\onlinecite{zeljkovic}] (see Fig. 1 of Ref. [\onlinecite{adou}]). This 
suggests that a possible source of the observed reconstruction is the 
onset of the structural bulk instability signaled at the surface. 
However, one should keep in mind that the observed symmetry breaking  
is limited to the surface only, since otherwise the band gap would 
strongly differ from that measured [\onlinecite{zeljkovic}].  

Second, breaking mirror symmetry leads to formation of surface dipoles 
[\onlinecite{nishi}]. This brings yet another question regarding the 
surface configuration: experiment reveals one direction of polarization 
only, while one would expect formation of surface domains with orthogonal 
orientations of polarization. 

Third, in the case of \pss\ the details of the reconstruction were 
directly displayed by the STM images [\onlinecite{zeljkovic}]. One can observe, 
however, an 
interesting feature exhibited in Fig. S3 of Ref. 
[\onlinecite{zeljkovic}]. Namely, the reconstruction pattern depends on 
the STM polarization voltage: at about $-250$ mV the symmetry-breaking 
mirror plane is (100), which changes to the (110) plane at higher 
voltages. This effect was not recognized by the authors, and is not 
reflected in our calculations. 
Finally, we cannot propose a source of this  contradiction. The above 
remarks are meant to underline the complexity of the issue, and not to 
question the quoted experimental data.

To study dispersion relations of surface states we use the tight binding
approach, with parameters obtained from the DFT calculations. As we
demonstrate, it is important to account for the position dependence 
of the TBA parameters (which differ for ions in the bulk and near 
the surface),
since otherwise the energies of surface states are erroneous. 
In particular, we analyze the transition from the NI to the TCI phase 
for the \pss. 
With the increasing Sn content the alloy undergoes the transition from
normal insulator  to the Weyl semimetal
phase. Because of the disorder induced band broadening, this phase
persists in a wide composition window 0.2 < x < 0.3, in which surface
states are characterized by atypical dispersion relations as well as 
semi-localization at the surface and, eventually, the TCI phase appears
for 0.3<~x~<~0.4.

\begin{acknowledgments}
This work was partially supported by National Science
Centre NCN (Poland) projects
UMO-2016/23/B/ST3/03725 (A{\L}), UMO-2017/27/B/ST3/02470
(A{\L}) as well as by the Foundation
for Polish Science through the IRA Programme co-financed
by EU within SG OP (TS). 
We thank R. Buczko for helpful discussions.
\end{acknowledgments}

\end{document}